\documentclass[12pt]{iopart}
\usepackage{graphicx}

\newcommand{\text}[1]{\mathrm{#1}}
\newcommand{\Hetrimer}{$^4$He$_3$}

\newcommand{\phig}{\phi_{\gamma}}
\newcommand{\phigp}{\phi_{\gamma'}}
\newcommand{\Eg}{E_{\gamma}}
\newcommand{\Egp}{E_{\gamma'}}

\newcommand{\DD}{D$_2$}
\newcommand{\oDD}{o-D$_2$}
\newcommand{\HH}{H$_2$}
\newcommand{\pHH}{p-H$_2$}
\newcommand{\HHDD}{H$_2$D$_2$}
\newcommand{\pHHoDD}{p-H$_2$--o-D$_2$}
\newcommand{\HHHH}{(\HH)$_2$}
\newcommand{\DDDD}{(\DD)$_2$}
\newcommand{\oDDDD}{(o-\DD)$_2$}

\newcommand{\pp}{\text{PP}}
\newcommand{\ttredpp}[1]{t_{#1}^\pp}
\newcommand{\tred}{t}
\newcommand{\Fggp}{F_{\gamma\gamma'}}
\newcommand{\taupp}[1]{\tau_{#1}^\pp}
\newcommand{\tauppn}{\tau^\pp}
\newcommand{\taudimggp}{\tau_{\gamma\gamma'}^\text{dim}}
\newcommand{\Inggp}{I_n^{\gamma\gamma'}}
\newcommand{\Inllp}{I_n^{ll'}}
\newcommand{\taudimlmlmp}{\tau_{lml'm'}^\text{dim}}
\newcommand{\Pillmp}{\Pi_{ll'}^{m'}}
\newcommand{\Watt}{W_{\text{att}}}

\renewcommand{\vec}[1]{\bi #1}  
\newcommand{\ket}[1]{\left| #1\right\rangle} 
\newcommand{\bra}[1]{\left\langle #1\right|}
\newcommand{\cc}[1]{#1^*}
\newcommand{\avg}[1]{\overline{#1}}

\newcommand{\trans}[2]{#1$\rightarrow$#2}

\begin{document}

\title{Inelastic Diffraction and Spectroscopy of Very Weakly Bound Clusters}
\author{Martin Stoll\dag\ and Thorsten K\"ohler\ddag}
\address{\dag\ Institut f\"ur Theoretische Physik, Universit\"at
  G\"ottingen, Bunsenstra{\ss}e 9, 37073 G\"ottingen, Germany}
\address{\ddag\ Clarendon Laboratory, Department of Physics, University
  of Oxford, Oxford OX1 3PU, United Kingdom}

\begin{abstract}
We study the coherent inelastic diffraction of very weakly bound
two body clusters from a material transmission grating. We show that
internal transitions of the clusters can lead to new separate peaks in 
the diffraction pattern whose angular positions determine the excitation
energies. Using a quantum mechanical approach to few body scattering 
theory we determine the relative peak intensities for the diffraction 
of the van der Waals dimers \DDDD\ and \HHDD. Based on the results for
these realistic examples we discuss the possible applications and experimental 
challenges of this coherent inelastic diffraction technique. 
\end{abstract}


\section{\label{sec:Intro}Introduction}

Since the first observation of Fresnel diffraction of atoms from a
single slit in 1969 \cite{LB_AJP37} diffraction of molecular beams
transmitted through material devices \cite{KSSP_PRL61,CM_PRL66} has
now become subject to intensive studies. New applications involve the
use of transmission gratings as quantum mechanical mass spectrometers
\cite{ST_SCIENCE266,CEHRSWP_PRL74}, the determination of atomic and
molecular electric dipole polarizabilities with interferometric
precision \cite{CEHRSWP_PRL74}, sensitive probing of the interaction
of atoms and molecules with solid surfaces
\cite{GSTHK_PRL83,BFGTHKSW_EPL59} as well as fundamental tests of the
quantum mechanical nature of complex molecules \cite{ANVKZA_N401}.
These recent developments all depend on micro-fabricated transmission
gratings whose periods have now become as small as 100 nm. These small
structures allow to scatter collimated molecular beams coherently. For
helium beams, e.g., far more than 10 diffraction orders have been
resolved whose intensities extend over more than four orders of
magnitude.

With this great sensitivity diffraction from material transmission
gratings can now be used for quantitative studies: Several recent
experiments have shown that classical wave optics is insufficient to
explain the relative magnitudes of the diffraction intensities
\cite{GSTHK_PRL83,GSTHKS_PRL85}. It turns out that a more general
approach based on quantum mechanical scattering theory is required to
include the van der Waals interaction between the atoms and the
grating bars and to account for the finite size of weakly bound
clusters in the beams. The highly precise diffraction experiments
together with this improved theoretical analysis have recently allowed
to determine the $C_3$ coefficients of the atom-surface van der Waals
interaction for a variety of atoms and molecules
\cite{GSTHK_PRL83,BFGTHKSW_EPL59} as well as the bond length of the
helium dimer \cite{GSTHKS_PRL85} which characterizes all low energy
binary scattering properties of helium.

These transmission grating diffraction studies only involve elastic
scattering. Atoms or molecules scattered from a solid surface can
undergo internal transitions. In reflection experiments from surface
lattices the coherent transfer between H$_2$ molecular rotational
levels was observed, e.g., in reference \cite{WYHLS_JCP83}.  The
possibility of coherent inelastic diffraction from transmission
gratings was discussed in reference \cite{HK_PRL84} for the particularly
interesting example of the excitation of the weakly bound helium
trimer \Hetrimer\ whose single excited state is believed to be an
Efimov state (see, e.g., reference \cite{ELG_PRA54}). These studies have
shown that, quite generally, an inelastic diffraction pattern exhibits
peaks that are separated from the strong elastic diffraction maxima
with an angular shift depending on the energy of the internal
transition and the grating period. The principle of inelastic
scattering from a material transmission grating, i.e.~the (de)excitation of
transmitted atoms or molecules, was demonstrated in
subsequent experiments using a fine structure transition of
meta-stable argon atoms\,\cite{BBD_EPJD17} and a vibrational
transition of meta-stable nitrogen dimers \cite{BBP_EPL56}. In these
experimental studies, however, the transmission grating served only to
multiply the intensity through the simultaneous scattering by all the
parallel surfaces of the bars. The inelastic diffraction peaks implied
by the periodicity of the grating have not yet been resolved.

Motivated by the recent progress in inelastic transmission scattering 
experiments in this article we study the coherent (de)excitation of very  
weakly bound two body clusters in the diffraction from a transmission grating.
Based on quantum mechanical few body scattering theory we discuss to 
which extent the small momentum transfers that the clusters experience when 
they pass through the slits can excite the low energy internal transitions. 
We show that the small angular shifts of the inelastic diffraction peaks 
can be well suited to determining the transition energies. 
The theory is applied 
to the realistic examples of the van der Waals clusters \DDDD\ and \HHDD. 
Their level spectra are well known both theoretically and experimentally 
\cite{D_JPB16,D_JPB22,McK_CJP52,McK_JCP92}.
Furthermore, experimental beam sources for these clusters are available.

\section{\label{sec:TMD}Diffraction of weakly bound dimers}

In this section we will briefly outline the few-body scattering approach
to the diffraction of weakly bound two-body systems 
\cite{HK_PRA57} that we will apply in this article. For simplicity,
we denote the two body systems as dimers although, 
quite generally, the constituents may be different atoms or tightly 
bound small 
molecules. These constituents are considered as point particles with 
masses $m_1$ and $m_2$ that interact with each other through a potential $V$. 
This potential is assumed to support shallow bound states $\phig$ with 
negative binding energy $\Eg$. 
\begin{figure}[htbp]
  \begin{center}
    \includegraphics[width=0.5\textwidth]{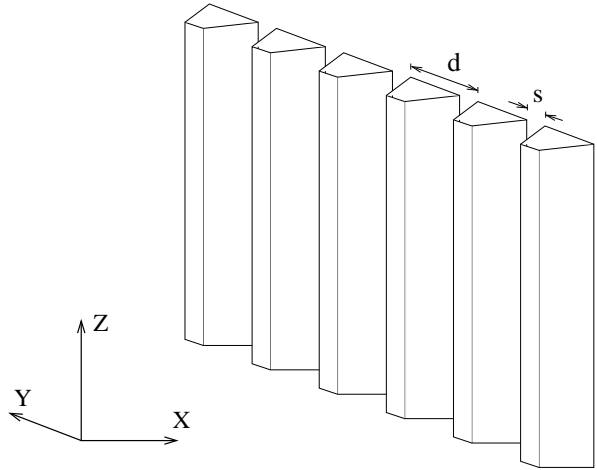}
    \caption{\label{fig:grating}Scheme of the material transmission grating 
      and the coordinate system used in this article. The bars are equally 
      spaced with a period $d$. The slit width of the grating is denoted 
      by $s$. The coordinates are chosen in such a way that the $z$ direction 
      is parallel to each bar and the grating is periodic in the $y$ 
      direction.}
  \end{center}
\end{figure}

The positions of the constituents are denoted by ${\vec x_1}$ and
${\vec x_2}$. The corresponding center of mass and relative coordinates
are ${\vec R}=(m_1{\vec x_1}+m_2{\vec x_2})/(m_1+m_2)$ and 
${\vec r}={\vec x_1}-{\vec x_2}$, respectively.
The diffracting object for the beam of dimers, i.e.~a material transmission 
grating in the applications in this article, is modeled by an external 
potential $W(\vec x_1,\vec x_2)$.
For the weakly bound dimers, under consideration, this two body potential 
is given, to an excellent approximation, by the sum of the potentials of 
each constituent, $W(\vec x_1,\vec x_2)=W_1(\vec x_1)+W_2(\vec x_2)$.
With the diffraction from a transmission grating in mind we idealize the
diffracting object in such a way that $W$ is invariant with respect to 
translations along the $z$ coordinate (see figure \ref{fig:grating}). 
This reduces the scattering problem, 
in the center of mass coordinates, effectively to two dimensions. 
The complete Hamiltonian is then given by
\begin{equation}
  \label{eq:fullhamilton}
  H=-\frac{\hbar^2{\vec \nabla}^2_{\vec R}}{2M}
  -\frac{\hbar^2{\vec \nabla}^2_{\vec r}}{2\mu}+
  V({\vec r})+W_1({\vec x}_1)+W_2({\vec x}_2),  
\end{equation}
where $M=m_1+m_2$ and $\mu=m_1 m_2/(m_1+m_2)$ denote the total and reduced 
mass, respectively.

A beam of dimers in the state $\phigp$ can be idealized as a
stationary energy state of the free center of mass motion. For an
incident center of mass momentum ${\vec P'}$ this state is given by
$\ket{\vec P',\phigp}=\ket{\vec P'}\ket{\phigp}$, where $\ket{\vec P'}$
is a plane wave. The total energy of a dimer in the beam then consists
of the kinetic energy and the binding energy, 
i.e.~$E'={\vec P'}^2/2M+\Egp$. Within the usual experimental range of 
(kinetic) beam energies internal excitations of the constituents 
should be negligible. As the dimers, under consideration, are weakly
bound, however, excitations of the relative motion of their constituents
as well as break-up are possible as long as the center of mass kinetic energy 
exceeds the dissociation threshold. We will study in this article the 
(de)excitation of dimers from the bound state $\phigp$ to $\phig$ at beam 
energies that are large in comparison to the break-up threshold. 
We will thus assume
\begin{equation}
  \label{eq:compareenergies}
  |E_0| \ll (P'^2_x+P'^2_y)/2M,
\end{equation}
where $E_0$ is the ground state binding energy of the dimer. 
We shall further presuppose the diffraction condition, i.e.~we assume that 
the de Broglie wave length of an incoming dimer, $\lambda'=2\pi\hbar/P'$, 
is by far smaller than all length scales set by the diffracting object.
The characteristic length scales of a transmission grating are the slit width 
$s$ and the width of the bars $d-s$ (see figure \ref{fig:grating}).
Under typical experimental conditions the ratios $s/\lambda'$ and 
$(d-s)/\lambda'$ are of the order of several hundreds or larger. 
The diffraction condition assures that the outgoing intensity is sharply 
peaked about the incident beam direction with a typical width of 
several mrad.  

The scattering matrix, $S$, maps the incoming state $\ket{\vec P',\phigp}$
onto the outgoing state, i.e.~the asymptotic state after the diffraction from 
the potential $W$. The $S$ matrix element that describes the (de)excitation of 
a dimer can be decomposed as
\begin{eqnarray}
  \nonumber
  \fl\bra{\vec P,\phig}S\ket{\vec P',\phigp}=
  \delta^{(3)}(\vec P-\vec P')\delta_{\gamma\gamma'}-2\pi\rmi\ \delta(E-E')
  \delta(P_z-P'_z)\\
  \label{eq:smatrixelement}
  \times\ t(P_x,P_y,\phig;P'_x,P'_y,\phigp),
\end{eqnarray}
where $E=\vec P^2/2M+\Eg$ is the energy of the outgoing dimer.
In reference \cite{HK_PRA57} the dimer transition amplitude
$t(P_x,P_y,\phig;P'_x,P'_y,\phigp)$ has been determined from a general
approach to few-body multi-channel scattering theory \cite{AGS}, by
means of perturbation theory.  According to this approach the dimer
transition amplitude can be expressed in terms of ``point particle''
transition amplitudes of the constituents:
\begin{equation}
  \label{eq:defttredpp}
  \ttredpp{i}(p_{ix},p_{iy}; \Delta p_{ix}, \Delta p_{iy})=
  \bra{p_{ix},p_{iy}}W_i\ket{p'_{ix},p'_{iy},+}_i.
\end{equation}
Here $p_{ik}'$ and $p_{ik}$, $i=1,2$, $k=x,y$ denote the momentum
components of a single constituent and $\Delta p_{ik}=p_{ik}-p'_{ik}$
the corresponding momentum transfers. Furthermore,
$\ket{p'_{ix},p'_{iy},+}_i$ is the stationary outgoing scattering
state that corresponds to the diffraction of constituent $i$ with
incident momentum $(p_{ix}',p_{iy}')$ from the potential $W_i$
\cite{HK_PRA57}.  From reference \cite{HK_PRA57} one then obtains
\begin{eqnarray}
  \nonumber
  \fl \tred(P_x,P_y,\phig;P'_x,P'_y,\phigp)=
  \ttredpp{1}\left(\frac{m_1}{M}P_x,\frac{m_1}{M}P_y;
    \Delta P_x,\Delta P_y\right)
  \Fggp\left(-\frac{m_2}{M}\Delta\vec P\right)\\
  \nonumber
  +\ \ttredpp{2}\left(\frac{m_2}{M}P_x,\frac{m_2}{M}P_y;
    \Delta P_x,\Delta P_y\right)
  \Fggp\left(\frac{m_1}{M}\Delta\vec P\right)\\
  \nonumber
  -\ \frac{2\pi\rmi M}{P_x}\int\!\!\rmd q_y\ 
  \ttredpp{1}\left(\frac{m_1}{M}P_x,\frac{m_1}{M}P_y; \Delta P_x,
    \frac{m_1}{M}\Delta P_y-q_y\right)\\
  \label{eq:treddim1}
  \times\ \ttredpp{2}\left(\frac{m_2}{M}P_x,\frac{m_2}{M}P_y; 0,
    \frac{m_2}{M}\Delta P_y+q_y\right)
  \Fggp\left(-\frac{m_2}{M}\Delta P_x,q_y,0\right),
\end{eqnarray}
where
\begin{equation}
  \label{eq:formfactor}
  \Fggp(\vec p)=\int\!\!\rmd^3r\ \exp(-\rmi\vec p\cdot\vec r/\hbar)\
  \cc{\phig}(\vec r)\phigp(\vec r)
\end{equation}
is usually referred to as the form factor of the dimer that corresponds
to the transition from $\phigp$ to $\phig$. 

In reference \cite{HK_PRA61} the transition amplitude (\ref{eq:defttredpp}) 
has been expressed in terms of a Fourier transform of a point particle 
transmission function $\taupp{i}$, $i=1,2$ (see Appendix):
\begin{eqnarray}
  \nonumber
  \fl \ttredpp{i}(p_{ix},p_{iy};\Delta p_{ix},\Delta p_{iy})=
    -\rmi\frac{p_{ix}}{(2\pi)^2 m_i\hbar}\\
  \label{eq:ttredpp}
  \times\ \int\!\!\rmd y\,
  \exp(-\rmi\Delta p_{iy} y/\hbar)
  \left[1-\taupp{i}(p'_{ix},p'_{iy};y)\right].
\end{eqnarray}
This representation is particularly useful in the description of
diffraction scattering close to the incident beam direction.  If, for
instance, the interaction of the constituents with the grating bars is
assumed to be purely repulsive $\taupp{i}$ recovers the grating
transmission function of classical optics and equation (\ref{eq:ttredpp})
becomes the classical Kirchhoff diffraction
amplitude\,\cite[chap. 8.5]{bornwolf}.

We shall use equation (\ref{eq:ttredpp})
to represent the dimer transition amplitude (\ref{eq:treddim1}) in terms
of a dimer transmission function: Inserting equation (\ref{eq:ttredpp}) into 
equation (\ref{eq:treddim1}) and performing the momentum integrals shows that,
under the diffraction condition and within the range of validity of 
assumption (\ref{eq:compareenergies}), some terms in 
equation (\ref{eq:treddim1}) 
cancel. The dimer transition amplitude then assumes the form:
\begin{eqnarray}
  \nonumber
  \fl \tred(P_x,P_y,\phig;P'_x,P'_y,\phigp)=
    -\rmi\frac{P_{x}}{(2\pi)^2 M\hbar}\\
  \label{eq:treddim2}
  \times\ \int\!\!\rmd Y\,
  \exp(-\rmi\Delta P_y Y/\hbar)
  \left[\delta_{\gamma\gamma'}-\taudimggp(P'_x,P'_y;Y)\right],
\end{eqnarray}
where the dimer transmission function is given by
\begin{eqnarray}
  \label{eq:taudimggp}
  \fl\taudimggp(P'_x,P'_y;Y)=\int\!\!\rmd^3r\ 
  \cc{\phig}(\vec r)\phigp(\vec r)\\
  \nonumber
  \times \
  \taupp{1}\left(\frac{m_1}{M}P'_x,\frac{m_1}{M}P'_y; 
    Y+\frac{m_2}{M}y\right)
  \taupp{2}\left(\frac{m_2}{M}P'_x,\frac{m_2}{M}P'_y; 
    Y-\frac{m_1}{M}y\right).
\end{eqnarray}
The coordinates ${\vec R}=(X,Y,Z)$ and ${\vec r}=(x,y,z)$ in 
equation (\ref{eq:taudimggp}) can be interpreted as center of mass and
relative coordinates, respectively. 

\section{\label{sec:DG}Diffraction from a Transmission Grating}

In this section we will analyze the kinematic diffraction phenomena
that are implied by the periodicity of a material diffraction grating
consisting of equally spaced bars with a period $d$ and a slit width
$s$ (see figure \ref{fig:grating}).  The grating potential of each
constituent is assumed to be strongly repulsive inside the bars with
an attractive part along their surfaces that accounts for the van der
Waals interaction of the constituents with the
material\,\cite{GSTHK_PRL83}.  The periodicity of the grating along
the $y$ axis reappears in the point particle transmission functions
$\taupp{i}$ and thus, by equation (\ref{eq:taudimggp}), the dimer
transmission function $\taudimggp$ is also periodic in the center of
mass coordinate $Y$. As a short calculation using
equation (\ref{eq:taudimggp}) shows, this periodicity implies the
conservation of the $y$ component of the center of mass momentum of a
dimer, up to reciprocal lattice vectors\footnote{This conservation
  law can also be deduced from general symmetry considerations.}:
\begin{equation}
  \label{eq:DeltaP2}
  \Delta P_y=n2\pi\hbar/d\ ,\quad n=0,\pm 1, \pm2,\ldots.
\end{equation}
Furthermore, the conservation of the total energy and the
translational invariance of the grating along the $z$ axis imply
\begin{equation}
\label{eq:erergcons}
(P_x'^2+P_y'^2)/2M+\Egp=(P_x^2+P_y^2)/2M+\Eg.
\end{equation}

These conservation laws determine, in turn, the angles of the
principal maxima of the diffraction intensity:
The angle of the $n$th order principal diffraction maximum, $\theta_n$, 
is given, in terms of the momentum $P_y$, by $P_y=P\sin\theta_n$.
In a similar way, $P'_y=P'\sin\theta'$ determines the angle of incidence of 
the beam in the $(x,y)$ plane perpendicular to each grating bar. 
Equations (\ref{eq:DeltaP2}) and (\ref{eq:erergcons}) then yield
\begin{equation}
  \label{eq:thetan}
  \sin\theta_n=\left(1-\frac{\Eg-\Egp}{P'^2/2M}\right)^{-1/2}
  \left[\sin\theta'+n\frac{2\pi\hbar}{P'd}\right].
\end{equation}
For elastic ($\Eg=\Egp$) diffraction this reproduces the well known
formula from wave optics. For inelastic ($\Eg\neq\Egp$) diffraction
$\theta_0$ is shifted to a larger (smaller) angle in the case of
excitation (de-excitation) and the spacing of the diffraction maxima
is increased (decreased). We note that
equation (\ref{eq:thetan}) only contains the (de)excitation energy 
$\Eg-\Egp$
but no further properties of the bound state wave functions. A physical
interpretation of equation (\ref{eq:thetan}), in terms of refraction of 
molecular beams, was given in reference \cite{HK_PRL84}. 
The break-up of a dimer
leads to diffuse scattering angles of the fragments and will not be 
considered further in this article. 

An experimental molecular beam exhibits a finite divergence which
leads to a broadening of the diffraction maxima. The intensity of a
diffraction order is usually obtained from the area under the
corresponding peak.  With this averaging procedure the diffraction
intensities become independent of the beam properties as long as the
peaks are resolved.  The diffraction intensity of the $n$th order
principal maximum for a transition between the bound states $\phigp$
and $\phig$ is then determined, in terms of the $S$ matrix
(\ref{eq:smatrixelement}), through
\begin{equation}
  \label{eq:Inppg0}
  \Inggp\propto\int_{C_n}\!\!\!\!\!\rmd^3P\ \left|
    \int\!\!\rmd^3P'\ \psi(\vec P')\ \bra{\vec P,\phig}S\ket{\vec P',\phigp}
  \right|^2.
\end{equation}
Here the momentum distribution $|\psi({\vec P'})|^2$ accounts for the 
beam divergence and $C_n$ denotes a small cone centered about the $n$th
order principal diffraction maximum. The integration over the cone determines
the area under the peak. 

To resolve the diffraction orders 
$\psi(\vec P')$ should be sharply peaked about an average momentum 
$\avg{\vec P'}$ which defines the incident direction as
well as the mean velocity of the beam of dimers. Under the assumption that the 
diffraction orders are resolved equations (\ref{eq:smatrixelement}), 
(\ref{eq:treddim2}) and (\ref{eq:Inppg0}) yield
\begin{equation}
  \label{eq:Inggp}
  \Inggp\propto\left|
    \frac1d\int_{-\frac d2}^\frac d2\!\!\rmd Y
    \exp(-\rmi 2\pi n Y/d)\
    \taudimggp\left(\avg{P'_x},\avg{P'_y};Y\right)
  \right|^2,
\end{equation}
where the integral extends only over one period of the grating. 
The intensities of the diffraction peaks are thus determined by 
the properties of the transmission function $\taudimggp$.

\section{\label{sec:examples}Examples: Diffraction of \DDDD\ and \HHDD}

We will now apply the above results to the van der Waals dimers \DDDD\ 
and \HHDD. To an excellent approximation the dimer binding potential
$V$ remains unchanged under isotopic substitution.  The different
small reduced masses of \DDDD\ and \HHDD, however, lead to pronounced
isotope shifts in their bound state spectra. The binding energies of
both species are of the order of only a few hundred $\mu$eV.  Along
with \HHHH\ these dimers were extensively studied, both theoretically
and experimentally, after finding evidence of \HHHH\ in astrophysical
planetary spectra.  According to their total nuclear spin one
distinguishes between the ortho (o) and para (p) modifications of \HH\ 
and \DD. All of them may combine to form dimers, leading to a rich
infrared fluorescence spectrum (see, e.g.,
references \cite{D_JPB16,D_JPB22,McK_CJP52,McK_JCP92}, and
references \cite{S_AA284,DJ_JCP112} for a recent overview of binding
potentials).

To facilitate the following discussion we will assume that, by passing
the gas through a catalytic converter or by a similar experimental
technique, the molecular beam contains only \oDD\ and \pHH. Both
molecules exhibit only even rotational angular momentum quantum
numbers $j$.  Furthermore, we shall assume that the nozzle temperature
of the beam apparatus is sufficiently low that the majority of the
\DD\ and \HH\ molecules are in their rotational ground states
$j=0$\,\cite{KDC_JCP82}.

\begin{figure}[htbp]
  \begin{center}
    \includegraphics[width=0.6\textwidth]{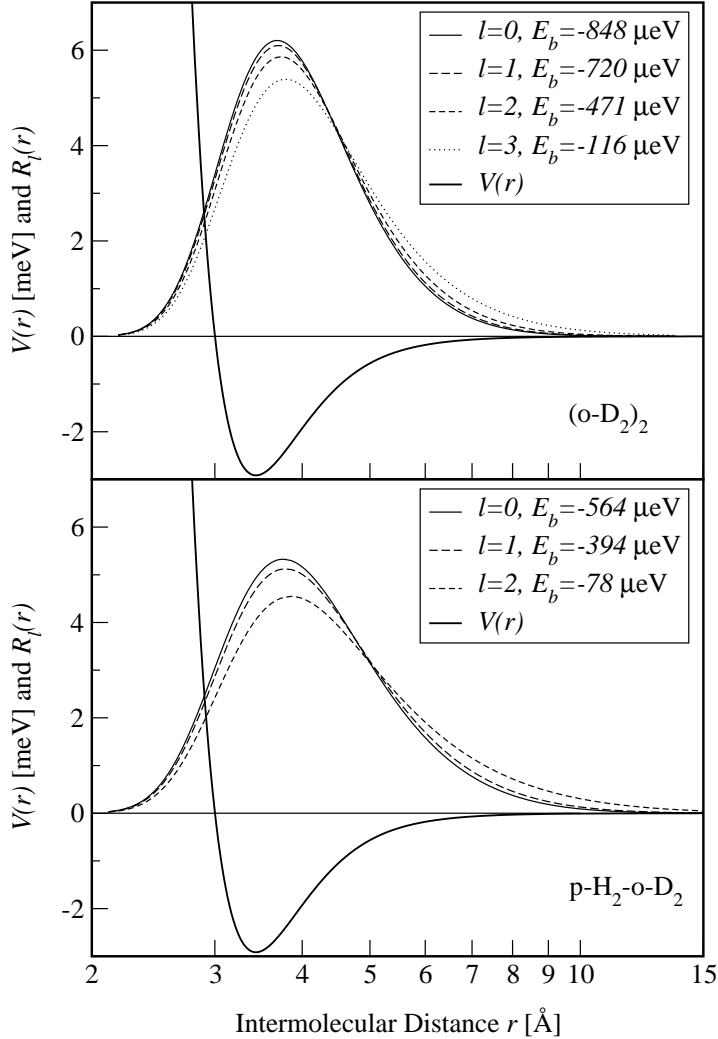}
    \caption{The radial wave functions
      $R_l(r)$ of \oDDDD\ (top) and \pHHoDD\ (bottom), where $l$
      denotes the end-over-end rotational quantum number. The
      corresponding binding energies are given in the legend. All
      bound states belong to the lowest vibrational modes.  Higher
      vibrational modes are unbound for both species. The radial coordinate
      is given on a logarithmic scale. The ordinates refer to the binding
      potential $V(r)$ from reference \cite{NWB_JCP81} which is shown for
      comparison.  \label{fig:wf}}
  \end{center}
\end{figure}

To determine the dimer transmission function from
equation (\ref{eq:taudimggp}) we first have to calculate the bound state
wave functions $\phig(\vec r)$ that describe the dimers \DDDD\ and
\HHDD.  To this end we have chosen the semi-empirical potential
$V({\vec r})$ by Buck \etal\,\cite{NWB_JCP81} which is given in
analytic form. Its anisotropic contributions will be neglected as we
assume the constituents to be mostly in spherically symmetric $j=0$
states. Then, as usual, the stationary two body Schr\"odinger equation
can be expanded in terms of spherical harmonics
$Y_l^m(\vartheta,\varphi)$ and one only needs to determine the radial
part. A numerical integration of the radial equation yields four
rotational bound states for \oDDDD\ and three for \pHHoDD, all belonging
to the lowest vibrational mode. As higher vibrational modes are unbound the 
radial bound state wave functions $R_l(r)$, as depicted in figure \ref{fig:wf},
are determined solely by their end-over-end rotational quantum number $l$.  
Within our approximations the binding energies of \oDDDD\ are in agreement 
with reference \cite{D_JPB22}.

The integration over the relative coordinate ${\vec r}=(x,y,z)$ in the
dimer transmission function (\ref{eq:taudimggp}) can be performed most
conveniently in spherical coordinates ($\int\rmd^3r=\int_0^\infty
r^2\rmd r \int_0^{2\pi}\rmd \varphi\int_0^\pi\sin\vartheta
\rmd\vartheta$).  By choosing $y=r\cos\vartheta$ the angular
integration over $\varphi$ becomes trivial and the substitution
$\alpha=\cos\vartheta$ yields
\begin{eqnarray}
  \nonumber
  \fl \taudimlmlmp(P'_x,P'_y;Y)=\delta_{mm'}
    \sqrt{2 l+1}\sqrt{2 l'+1} \int_0^\infty\!\!\!\! r^2 
    \rmd r\ R_l(r)R_{l'}(r)
    \int_{-1}^{1}\!\!\rmd \alpha\ \Pillmp(\alpha)\\
  \label{eq:taudimlmlmp}
  \times\ 
  \taupp{1}\left(\frac{m_1}{M}P'_x,\frac{m_1}{M}P'_y; 
    Y+\frac{m_2}{M}\alpha r\right)
  \taupp{2}\left(\frac{m_2}{M}P'_x,\frac{m_2}{M}P'_y; 
    Y-\frac{m_1}{M}\alpha r\right).
\end{eqnarray}
Here the function $\Pillmp(\alpha)$ has been introduced as
\begin{equation}
  \label{eq:Pillm}
  \Pillmp(\alpha)=\frac12
  \sqrt{\frac{(l-|m'|)!\ (l'-|m'|)!}{(l+|m'|)!\ (l'+|m'|)!}}\
  P_l^{|m'|}(\alpha)P_{l'}^{|m'|}(\alpha),
\end{equation}
where $P_l^m$ are the associated Legendre polynomials. $\Pillmp(\alpha)$
is symmetric in the indices $l$ and $l'$. For identical constituents the 
symmetry properties of the associated Legendre polynomials with respect to 
reflections at the origin imply a parity conservation selection rule for the
inelastic diffraction intensity:
\begin{equation}
  \label{eq:selectionrule}
  l'+l=\text{even.}
\end{equation}
This selection rule can be obtained from equation (\ref{eq:taudimlmlmp}) 
when one
inserts $m_1=m_2=M/2$ but may also be derived directly from general 
symmetry properties of the Hamiltonian (\ref{eq:fullhamilton}).

We assume the dimer bound states in the incident beam to be populated
in accordance with the thermal equilibrium weight factors
$p_{l'}=(2l'+1)\exp(-E_{l'}/k_{\rm B}T_b)$ where $T_b$ is the
translational beam temperature and $k_{\rm B}$ is the Boltzmann
constant. Inserting equation (\ref{eq:taudimlmlmp}) into
equation (\ref{eq:Inggp}) and summing over $m'$ we find that the $n$th
order diffraction intensity for a transition \trans{$l'$}{$l$} is
given by
\begin{eqnarray}
  \nonumber
  \fl \Inllp\propto (2l+1)\ p_{l'} \sum_{m'}
  \Bigg|
    \frac1d\int_{-\frac d2}^\frac d2\!\!\rmd Y
    \exp(-\rmi 2\pi n Y/d)
    \int_0^\infty\!\!\!\! r^2\rmd r\ R_l(r)R_{l'}(r)
    \int_{-1}^{1}\!\!\rmd \alpha\ \Pillmp(\alpha)\\
  \label{eq:Inllp}
  \!\!\!\!\!\!\!\!\!\!\!\!\!\!\!\!\!\!\!\!\!\!\!\!\!
  \times\
  \taupp{1}\left(\frac{m_1}{M}P'_x,\frac{m_1}{M}P'_y; 
    Y+\frac{m_2}{M}\alpha r\right)
  \taupp{2}\left(\frac{m_2}{M}P'_x,\frac{m_2}{M}P'_y; 
      Y-\frac{m_1}{M}\alpha r\right)
  \Bigg|^2,
\end{eqnarray}
where, due to the factor $\delta_{mm'}$ in equation (\ref{eq:taudimlmlmp}), 
only the orientation quantum numbers $m'$ between $-\min(l,l')$
and $\min(l,l')$ contribute to the sum. 

By considering \DD\ and \HH\ as point particles we have neglected
transitions between rotational states of the constituents. Due to their 
small moments of inertia, however, the
$j$=\trans02 transition energies are comparatively large, i.e.~22 meV for 
\oDD\ and 44 meV for \pHH\ \cite{KDC_JCP82}. By keeping the kinetic
energy in the beam below these thresholds excitation of higher $j$ states
can, therefore, be ruled out.

The diffraction patterns in the following two subsections (figures
\ref{fig:DDDD} and \ref{fig:HHDD}) were calculated with experimentally
realizable parameters in mind. The mean beam velocity was chosen as
$v'=\avg{P'}/M=500$ ms$^{-1}$ with a velocity spread of $\Delta
v'/v'=8\%$~\cite{W_RGD79,GSTHK_PRL83}. This corresponds to a kinetic
energy of 5.2 meV for \DD\ (2.6 meV for \HH), and a translational beam
temperature of $T_b\approx0.4$ K in a pure \DD-beam. We have
considered diffraction from a typical silicon nitride grating, as
characterized experimentally in reference \cite{GSTHK_PRL83}, with a
grating period of $d=100$ nm, a slit width of $s=60$ nm, and a wedge
angle of the grating bars of $\beta=5^\circ$. The attractive van der
Waals interaction between the grating and the constituents is
accounted for in the point particle transmission functions. Its $C_3$
coefficient (see Appendix) has been determined experimentally for
silicon nitride and \DD\ as $C_3=0.32$ meV nm$^3$ \cite{GSTHK_PRL83}.
We are not aware of a corresponding experimental value for \HH\ which
we therefore estimated: According to Hoinkes' empirical rule
\cite{H_RMP52,GSTHK_PRL83,BFGTHKSW_EPL59} $C_3$ is, quite generally,
proportional to the static electric dipole polarizability of the atom
or molecule. The polarizabilities of \HH\ and \DD\ differ by about 1\%
only. We thus assumed the same $C_3$ for both \DD\ and \HH.

In a typical experiment the diffraction maxima are broadened due to the beam
divergence and the width of the detector slit. The finite velocity spread of
a beam leads to an additional broadening which increases with the diffraction 
order $n$. These mechanisms are all specific to the experimental setup. 
We, therefore, use an empirical model to include the broadening. Previous 
experimental studies \cite{GSTHK_PRL83,GSTHKS_PRL85} have shown that the 
maxima can be represented, to an excellent approximation, by Gaussians of the 
form
\begin{displaymath}
  \frac{I}{\sqrt{\pi}w_n}
  \exp\left(-{\left(\theta-\theta_n\right)^2}/{w_n^2}\right)
\end{displaymath}
whose widths $w_n$ depend on the diffraction order $n$ as
\begin{displaymath}
  w_n=w_0\sqrt{1+\left(\frac{\Delta w}{w_0}n\right)^2}.
\end{displaymath}
Here $w_0$ is the angular width of the zeroth order diffraction
maximum and $\Delta w$ accounts for the velocity spread. For the beams
in the present applications we have chosen $w_0=3\times10^{-3}$
degrees and $\Delta w=7\times10^{-4}$ degrees. Each diffraction
maximum is thus represented by a Gaussian of this kind with an area
$I$ given by equation (\ref{eq:Inllp}).  The angle of incidence $\theta'$
was chosen in such a way that the zeroth order principal diffraction
maxima of all inelastic transitions do not overlap with the zeroth
order elastic peak.

\subsection{Diffraction of \oDDDD}

In the case of \oDDDD\ both constituents are identical and the
selection rule (\ref{eq:selectionrule}) applies. Table
\ref{tab:transDD} shows the allowed transitions and the corresponding
excitation energies.
\begin{table}[htbp]
  \caption{\label{tab:transDD}Allowed excitations \trans{$l'$}{$l$} for \oDDDD\
    and transition energies for the dimer binding potential from
    reference \cite{NWB_JCP81}. Also allowed are the reverse processes
    (de-excitation) with negative transition energies.}
  \begin{indented}
  \item[]
    \begin{tabular}{cc}
      \br
      \trans{$l'$}{$l$} & $E_{l}-E_{l'}$ [$\mu$eV]\\
      \mr
      elastic ($l'=l$) & 0\\
      \trans02 & 377\\
      \trans13 & 604\\
      \br
    \end{tabular}
  \end{indented}
\end{table}
Due to the low translational beam temperature of $T_b\approx0.4$ K most dimers 
are in the rotational ground state $l'=0$. The population of the $l'=1$
state is only of the order of 7\%, and the initial populations of
higher excited states are negligible.

\begin{figure}[htbp]
  \begin{center}
    \includegraphics[width=0.6\textwidth]{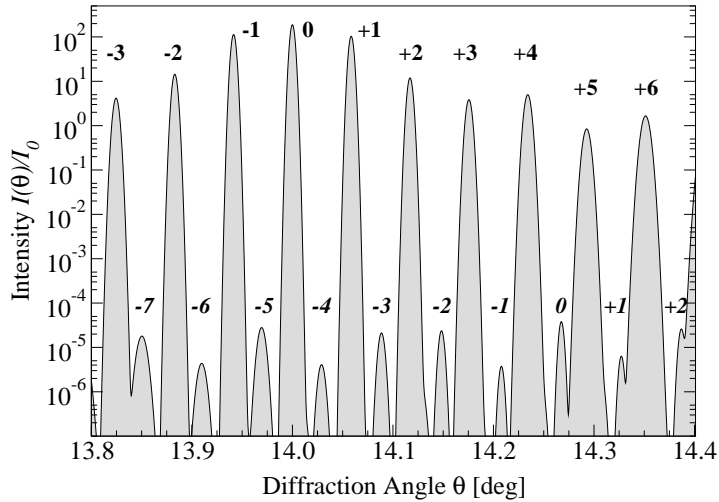}
    \caption{Diffraction pattern of a pure \oDDDD\ beam including all
      allowed transitions \trans{$l'$}{$l$}. The angle of incidence is 
      $\theta'=14^\circ$.
      The intense diffraction maxima are due to elastic scattering.
      The italic labels indicate the inelastic \trans02 diffraction orders.
      Other inelastic transitions are masked or too weak to be
      resolved. The diffraction pattern is normalized to the total
      elastic zeroth order intensity
      $I_0=I_0^{00}+I_0^{11}+I_0^{22}+I_0^{33}$.\label{fig:DDDD}}
  \end{center}
\end{figure}

Figure \ref{fig:DDDD} shows a \oDDDD\ diffraction pattern with an angle
of incidence of $\theta'=14^\circ$. The intense maxima correspond to
elastic scattering whereas the weak maxima result from the inelastic
\trans02 transition. The even weaker \trans13 maxima are too close to
the elastic diffraction peaks to be resolved. According to
equation (\ref{eq:thetan}) the spacing of the \trans02 maxima and the shift
of the zeroth order with respect to the incident beam direction
determine the excitation energy $E_2-E_0=377\mu$eV. The hierarchy of
the inelastic maxima depends on the specific interaction with the
grating bars.

\subsection{Diffraction of \pHHoDD}

In the case of \pHHoDD\ the selection rule (\ref{eq:selectionrule}) does
not apply, allowing all transitions between the three bound states.
Their excitation energies are shown in table \ref{tab:transHHDD}.
\begin{table}[htbp]
  \caption{\label{tab:transHHDD}Allowed excitations \trans{$l'$}{$l$} 
    for \pHHoDD\ and
    transition energies for the dimer binding potential from
    reference \cite{NWB_JCP81}. Also allowed are the reverse processes
    (de-excitation) with negative transition energies.}
  \begin{indented}
    \item[]
    \begin{tabular}{cc}
      \br
      \trans{$l'$}{$l$} & $E_{l}-E_{l'}$ [$\mu$eV]\\
      \mr
      elastic ($l'=l$) & 0\\
      \trans01 & 170\\
      \trans02 & 486\\
      \trans12 & 316\\
      \br
    \end{tabular}
  \end{indented}
\end{table}
Again, most dimers are in the rotational ground state $l'=0$. About
2\% of the dimers are in the $l'=1$ state, and the initial population of
the $l'=2$ state is negligible.

\begin{figure}[htbp]
  \begin{center}
    \includegraphics[width=0.6\textwidth]{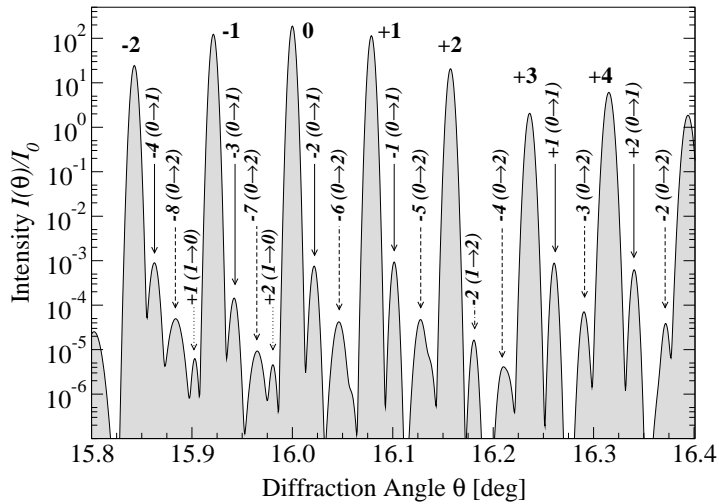}
    \caption{Diffraction pattern of a pure \pHHoDD\ beam
      including all allowed transitions. The angle of incidence is
      $\theta'=16^\circ$. The intense diffraction maxima are due to
      elastic transitions. The italic labels indicate the resolvable 
      inelastic diffraction orders. The diffraction pattern is normalized to
      the total elastic zeroth order intensity
      $I_0=I_0^{00}+I_0^{11}+I_0^{22}$.\label{fig:HHDD}}
  \end{center}
\end{figure}

Figure \ref{fig:HHDD} shows a \pHHoDD\ diffraction pattern with an angle
of incidence of $\theta'=16^\circ$. Besides the intense elastic maxima
the inelastic transitions \trans01 and \trans02 are resolved. Their
spacings determine the transition energies of $E_1-E_0=170\mu$eV and
$E_2-E_0=486\mu$eV, respectively. The \trans01 diffraction maxima are,
in general, about an order of magnitude more intense than the \trans02
ones. It is interesting to note, however, that the \trans01 zeroth
order is strongly suppressed because the point particle transmission
functions of \HH\ and \DD\ are almost equal and almost symmetric. The
zeroth order vanishes exactly in the hypothetical limit that the van
der Waals interaction of the constituents with the grating vanishes.
Also visible, but weaker because of the small initial $l'=1$
population, are some maxima of \trans10 and \trans12 transitions.

\section{Conclusions}
We have shown in this article how inelastic diffraction from a 
transmission grating may be used to study energy spectra
of very weakly bound two body clusters. The transition energies 
are determined by angular shifts of the inelastic diffraction
peaks in comparison to the strong elastic peaks. The emergence of these
peaks is implied by the periodicity of the grating. We have determined
their relative intensities from a quantum mechanical few body scattering
approach for the examples of the van der Waals dimers \DDDD\ 
and \HHDD.

For these realistic examples the angles of the inelastic diffraction peaks 
can be resolved in present day experiments while their small intensities 
would require an additional gain in sensitivity. This gain may be achieved, 
for instance, by increasing the detection efficiency and, thereby, keeping 
the background detection rate at a constant level. 

Given a suitable detection efficiency the inelastic diffraction
technique described in this article should be applicable also to those
very weakly bound clusters that are not easily excited by laser light.
A prominent example of this kind, the helium trimer, was discussed in
reference \cite{HK_PRL84}.  Inelastic diffraction from a transmission
grating may also be used to separate beams of clusters in particular
internal states from the incoming beam.

\ack
We would like to thank R.~Br{\"u}hl, G.~C.~Hegerfeldt and
J.~P.~Toennies for interesting discussions. This research was supported
by the Alexander von Humboldt Foundation and the United Kingdom EPSRC.

\appendix
\section*{Appendix. The Point Particle Transmission Function}
\begin{figure}[htbp]
  \begin{center}
    \includegraphics[width=0.5\textwidth]{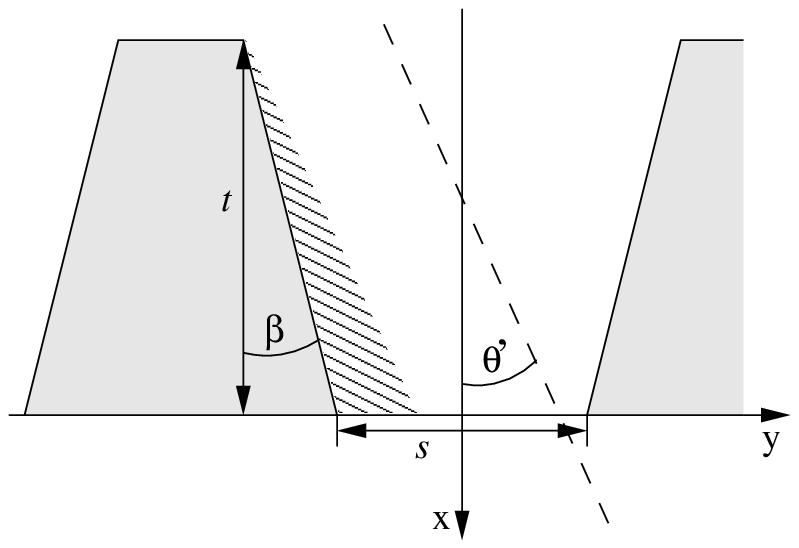}
    \caption{The trapezoidal shape of the grating bars with thickness
      $t$ and wedge angle $\beta$, and the direction of the incident
      beam (dashed line). For $\theta'>\beta$ a portion of the slit is
      shadowed (hatched region).}
    \label{fig:taupp}
  \end{center}
\end{figure}

\setcounter{section}{1}
In reference \cite{HK_PRA61} the point 
particle transmission function was derived 
in terms of an exact scattering solution of the two-dimensional stationary 
Schr\"odinger equation as (cf.~figure \ref{fig:taupp}):
\begin{displaymath}
  \tauppn(p'_x,p'_y;y)=\exp(-\rmi p'_y y/\hbar)\varphi(0,y).
\end{displaymath}
Here this scattering solution $\varphi(x,y)$ assumes, at large distances 
from the grating, the asymptotic form of a coherent superposition of the 
incoming plane wave $\exp[\rmi(p'_x x+p'_y y)/\hbar]$ and an outgoing 
cylindrical wave. In this appendix we will explain the methods we have used 
to determine $\tauppn(y)$. We shall focus on off normal incidence of the 
incoming beam ($\theta' \neq 0$) as the case of normal incidence on the 
transmission grating was discussed in detail in 
references \cite{HK_PRA61,GSTHK_PRL83}.
The coordinates as well as the typical trapezoidal
shape of the grating bars \cite{GSTHK_PRL83} are depicted schematically in 
figure \ref{fig:taupp}. As the grating bars are assumed to reflect those atoms 
that touch the bar walls the wave function $\varphi(x,y)$ vanishes at the 
surface of the bars. This implies $\tauppn(y)=0$ for the coordinates $y$ at 
the rear of a grating bar. The point particle transmission function 
$\tauppn(y)$ in the slits can be determined in accordance with the standard 
eikonal approximation \cite{joachain}: 
When, at off normal incidence, $\theta'$ exceeds the wedge 
angle $\beta$ a portion $t(\tan\theta'-\tan\beta)$ of each slit is shaded 
by the adjacent grating bar (see figure \ref{fig:taupp}). 
In this case $\tauppn(y)$ also vanishes in the shadow region.
For the coordinates $y$ inside the remaining open fraction of a slit
the point particle transmission function assumes the form
$\tauppn(y)=\exp[\rmi\Phi(y)]$.  The phase shift $\Phi(y)$ accounts for
the attractive van der Waals interaction between the atom or molecule
and the grating bars. It is determined by
\begin{equation}
  \Phi(y)=-\frac{1}{\hbar v'}\int\!\!\rmd u\ \Watt(u),
  \label{eq:phaseshift}
\end{equation}
where $v'=\sqrt{p'^2_x + p'^2_y}/m$ is the velocity of the incoming atom or 
molecule and the integration is performed along a straight line in the 
direction of the beam 
(dashed line in figure \ref{fig:taupp}).

The long range van der Waals potential between a pair of atoms (or molecules)
separated by a distance $R$ is proportional to $-R^{-6}$. To determine the
functional form of the attractive interaction between an atom or molecule and 
the grating one may use a pairwise summation of the interactions between the 
probe particle and the constituents of the solid. For an idealized 
flat surface, i.e.~when the solid covers the half
space, the pairwise summation implies the well known functional form 
$\Watt=-C_3/L^3$, where $L$ denotes the distance between the probe particle 
and the surface. The coefficient $C_3$ depends on the static
electric dipole polarizability of the atom or molecule and the material of 
the solid
\cite{GSTHK_PRL83,H_RMP52}. For \DD\ and a silicon nitride grating $C_3$
is known experimentally \cite{GSTHK_PRL83}. For \HH\ $C_3$ has been
estimated as explained in Section \ref{sec:examples}.

For a grating bar with the geometrical form in figure \ref{fig:taupp}
the attractive interaction $\Watt$ does not assume a simple form. By
interchanging the pairwise summation with the integral in equation
(\ref{eq:phaseshift}), however, the phase shift may be determined
analytically. Numerical studies have shown that only those bars
contribute significantly to $\Watt$ that confine the actual slit.  At
off normal incidence with $\theta'>\beta$ the phase shift is then
given by
\begin{equation}
  \Phi(y)=\frac{C_3}{2\hbar v' (\cos\theta')^4}
  \left[\frac{\xi_{11}^{-2}-\xi_{12}^{-2}}{\tan\theta'+\tan\beta}+
    \frac{\xi_{21}^{-2}-\xi_{22}^{-2}}{\tan\theta'-\tan\beta}\right],
  \label{eq:phaseshift2}
\end{equation}
where $\xi_{11}=s/2-y$, $\xi_{12}=s/2+t(\tan\beta+\tan\theta')-y$,
$\xi_{21}=s/2+t(\tan\beta-\tan\theta')+y$ and $\xi_{22}=s/2+y$.


\section*{References}
\begin{thebibliography}{10}

\bibitem{LB_AJP37}
Leavitt J~A and Bills F~A 1969 {\it Am.~J.~Phys.\/} {\bf 37} 905

\bibitem{KSSP_PRL61}
Keith D~W, Schattenburg M~L, Smith H~I and Pritchard D~E 1988 {\it
  Phys.~Rev.~Lett.\/} {\bf 61} 1580

\bibitem{CM_PRL66}
Carnal O and Mlynek J 1991 {\it Phys.~Rev.~Lett.\/} {\bf 66} 2689

\bibitem{ST_SCIENCE266}
Sch{\"o}llkopf W and Toennies J~P 1994 {\it Science\/} {\bf 266} 1345

\bibitem{CEHRSWP_PRL74}
Chapman M~S, Ekstrom C~R, Hammond T~D, Rubenstein R~A, Schmiedmayer J, Wehinger
  S and Pritchard D~E 1995 {\it Phys.~Rev.~Lett.\/} {\bf 74} 4783

\bibitem{GSTHK_PRL83}
Grisenti R~E, Sch{\"o}llkopf W, Toennies J~P, Hegerfeldt G~C and K{\"o}hler T
  1999 {\it Phys.~Rev.~Lett.\/} {\bf 83} 1755

\bibitem{BFGTHKSW_EPL59}
Br{\"u}hl R, Fouquet P, Grisenti R~E, Toennies J~P, Hegerfeldt G~C, K{\"o}hler
  T, Stoll M and Walter C 2002 {\it Europhys.~Lett.\/} {\bf 59} 357

\bibitem{ANVKZA_N401}
Arndt M, Nairz O, Vos-Andreae J, Keller C, van~der Zouv G and Zeilinger A 1999
  {\it Nature (London)\/} {\bf 401} 680

\bibitem{GSTHKS_PRL85}
Grisenti R~E, Sch{\"o}llkopf W, Toennies J~P, Hegerfeldt G~C, K{\"o}hler T and
  Stoll M 2000 {\it Phys.~Rev.~Lett.\/} {\bf 85} 2284

\bibitem{WYHLS_JCP83}
Whaley K~B, Yu C, Hogg C~S, Light J~C and Sibener S~J 1985 {\it
  J.~Chem.~Phys.\/} {\bf 83} 4235

\bibitem{HK_PRL84}
Hegerfeldt G~C and K{\"o}hler T 2000 {\it Phys.~Rev.~Lett.\/} {\bf 84} 3215

\bibitem{ELG_PRA54}
Esry B~D, Lin C~D and Greene C~H 1996 {\it Phys.~Rev.~A\/} {\bf 54} 394

\bibitem{BBD_EPJD17}
Boustimi M, Baudon J, Ducloy M, Reinhardt J, Perales F, Mainos C, Bocvarski V
  and Robert J 2001 {\it Eur.~Phys.~J.~D\/} {\bf 17} 141

\bibitem{BBP_EPL56}
Boustimi M, Baudon J, Pirani F, Ducloy M, Reinhardt J, Perales F, Mainos C,
  Bocvarski V and Robert J 2001 {\it Europhys.~Lett.\/} {\bf 56} 644

\bibitem{D_JPB16}
Danby G 1983 {\it J.~Phys.~B\/} {\bf 16} 3411

\bibitem{D_JPB22}
Danby G 1989 {\it J.~Phys.~B\/} {\bf 22} 1785

\bibitem{McK_CJP52}
McKellar A~R~W and Welsh H~L 1974 {\it Can.~J.~Phys.\/} {\bf 52} 1082

\bibitem{McK_JCP92}
McKellar A~R~W 1990 {\it J.~Chem.~Phys.\/} {\bf 92} 3261

\bibitem{HK_PRA57}
Hegerfeldt G~C and K{\"o}hler T 1998 {\it Phys.~Rev.~A\/} {\bf 57} 2021

\bibitem{AGS}
Alt E~O, Grassberger P and Sandhas W 1967 {\it Nucl.~Phys.~B\/} {\bf 2} 167

\bibitem{HK_PRA61}
Hegerfeldt G~C and K{\"o}hler T 2000 {\it Phys.~Rev.~A\/} {\bf 61} 23606

\bibitem{bornwolf}
Born M and Wolf E 1959 {\it Priciples of Optics\/} (Pergamon Press)

\bibitem{S_AA284}
Schaefer J 1994 {\it Astron.~Astrophys.\/} {\bf 284} 1015

\bibitem{DJ_JCP112}
Diep P and Johnson J~K 2000 {\it J.~Chem.~Phys.\/} {\bf 112} 4465

\bibitem{KDC_JCP82}
Kern K, David R and Comsa G 1985 {\it J.~Chem.~Phys.\/} {\bf 82} 5673

\bibitem{NWB_JCP81}
Norman M~J, Watts R~O and Buck U 1984 {\it J.~Chem.~Phys.\/} {\bf 81} 3500

\bibitem{W_RGD79}
Winkelmann K 1979 in {\it Rarefied Gas Dynamics\/}, ed. R~Camparque (CEA,
  Paris) vol.~2 p. 899

\bibitem{H_RMP52}
Hoinkes H 1980 {\it Rev.~Mod.~Phys.\/} {\bf 52} 933

\bibitem{joachain}
Joachain C~J 1975 {\it Quantum Collision Theory\/} (North-Holland Physics
  Publishing)

\end{thebibliography}
\end{document}